\documentclass[reprint, amsmath,amssymb, aps,]{revtex4-2}

\usepackage{graphicx}
\usepackage{xcolor}
\usepackage{dcolumn}
\usepackage{bm}
\usepackage{hyperref}

\begin{document}

\title{Scanning force sensing at $\mu$m-distances from a conductive surface with nanospheres in an optical lattice} 

\author{Cris Montoya$^1$, Eduardo Alejandro$^1$, William Eom$^1$, Daniel Grass$^1$, Nicolas Clarisse$^1$, Apryl Witherspoon$^2$, and Andrew A. Geraci$^1$ }
    \affiliation{$^1$Center for Fundamental Physics, Department of Physics and Astronomy, Northwestern University, Evanston, Illinois 60208, USA}
    
    \affiliation{$^2$Department of Physics, University of Nevada, Reno, NV 89557}

    \email[Correspondence email address: ]{andrew.geraci@northwestern.edu}

\date{\today}

\begin{abstract}
The center-of-mass motion of optically trapped dielectric nanoparticles in vacuum is extremely well-decoupled from its environment, making a powerful tool for measurements of feeble sub-attonewton forces. We demonstrate a method to trap and manuever nanoparticles in an optical standing wave potential formed by retro-reflecting a laser beam from a metallic mirror surface. We can reliably position a $\sim 170$ nm diameter silica nanoparticle at distances of a few hundred nanometers to tens of microns from the surface of a gold-coated silicon mirror by transferring it from a single-beam tweezer trap into the standing wave potential. We can further scan the two dimensional space parallel to the mirror surface by using a piezo-driven mirror.  
This method enables three-dimensional scanning force sensing near surfaces using optically trapped nanoparticles, promising for high-sensitivity scanning force microscopy, tests of the Casimir effect, and tests of the gravitational inverse square law at micron scales.
\end{abstract}

\maketitle

\section{Introduction} \label{sec:intro}
    Optical trapping of dielectric nanoparticles in vacuum has provided the optomechanics community with a tool for precise measurements at room temperature of force \cite{Hempston_ForceSensing_2017,Ranjit_zeptonewton_2016,Rodenburg_quantumModel_2016}, acceleration \cite{Monteiro_accelerationSensitivity_2017}, torque \cite{ahn_ultrasensitiveTorque_2020}, and charge \cite{Moore_Millicharged_2014} due to excellent decoupling from the environment, corresponding to high mechanical quality factors exceeding $10^7$ \cite{novotny2012}. This technology enables studies of fundamental physics including the Casimir effect \cite{Casimir,geraci_shortRange_2010}, searches for novel gravitational-like forces at sub-millimeter range \cite{geraci_shortRange_2010}, and gravitational waves \cite{GWprl,LSDpaper}, as well as development of sensitive inertial sensors \cite{pratCamps_diamagneticallyLevitated_2017}, or sensors for electric fields \cite{Ranjit_zeptonewton_2016,Moore_Millicharged_2014}. Levitated optomechanical systems have also recently proven to be useful for observing quantum behavior in the motional degrees of freedom in mechanical systems \cite{chang2009,Oriol2011,coherentscattering, aspelmeyercavity,aspelmeyergroundstate,magrini2020optimal,novotny2020,kamba2021recoillimited}, and may lead to the development of new tools for quantum information science perhaps especially when coupled to other quantum systems \cite{sympcool}. Ultracold nanoparticles can be used as a source for matter wave inteferometry \cite{Ulbricht:2014}, leading to tests of quantum phenomena at macroscopic scales \cite{Oriol2011other} and precision sensing \cite{andyhart2015}.
    
    In many experiments, such as those intended to measure the Casimir effect \cite{casimirpolder} or search for ``fifth-forces'' that may modify Newton's inverse square law of gravitation at short range \cite{GiudiceDimopoulos,add}, levitation of the particle near a surface is of particular interest. Having the capability to scan the force sensor in three dimensions in the vicinity of the surface enables mapping out surface forces from permanent and induced electric dipole moments and local patch potentials \cite{patch} in conductors, which can aid in understanding spurious backgrounds that often can limit such fundamental physics tests \cite{Kapner2007, Geraci2008, Chen2016}. 
    
    In this paper, we describe a protocol for trapping and maneuvering a $\sim 170$ nm silica nanosphere near a metallic mirror surface. Initially, the particle is trapped in a vacuum chamber using a single-beam optical tweezer. The trap is then changed into an optical lattice by introducing the reflective surface on a piezo-driven in-vacuum translation stage, and the particle is transferred in an anti-node of the standing wave which provides a localized confinement near the surface. By adjusting the divergence of the input trapping beam, the separation of the particle from the surface can be controlled at the $\sim \mu$m level, and by adjusting an in-vacuum piezo-driven mirror mount, the particle can be scanned in the two dimensional space parallel to the surface. 
    We examine the force sensitivity of the detector in the vicinity of the surface and show that at the vacuum level studied, there is no significant degradation of force sensitivity as the particle approaches the surface,  corresponding to attonewton scale forces. Thus our method is promising for ultra-sensitive three-dimensional scanning force microscopy \cite{quate,saridreview}. By functionalizing the particle with charge or a magnetic moment, in principle scanning electrometry \cite{electrometer} and scanning magnetic resonance force microscopy \cite{Rugar2004} are enabled via a similar approach. 

\section{Experimental Setup} \label{sec:setup}
    
\begin{figure}
	\begin{center}
		\includegraphics[width=1.0\columnwidth]{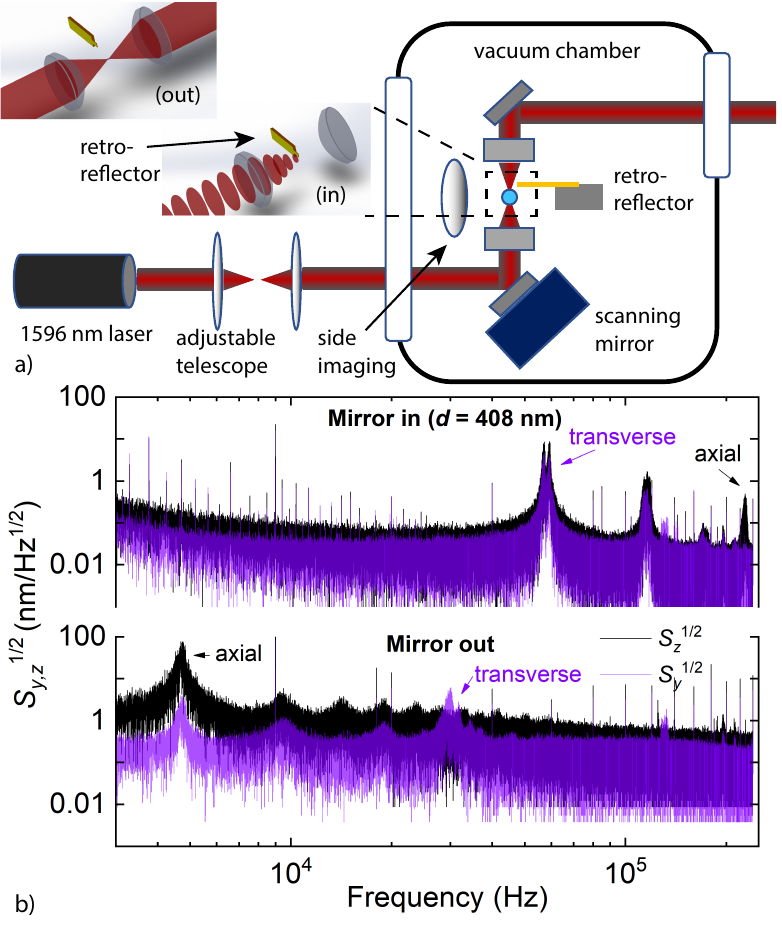}
	\end{center}
    \caption[Setup]{(a) Schematic of optical trapping apparatus with movable retro-reflecting and scanning mirrors for transferring nanoparticles from an optical tweezer trap to a standing wave trap and maneuvering them near a conducting surface. (inset) A 1596 nm laser beam is focused to create an optical tweezer where a 177 nm silica sphere is trapped. A gold-coated silicon tip retro-reflects the laser beam to create a standing wave where the silica nanosphere remains trapped. (b) Power spectral density as observed through imaging the scattered light from the side imaging lens on a quadrant-photodetector for a trapped nanosphere at a pressure of 100 mTorr in the tweezer (lower) and in the standing wave trap (upper) at a distance of $408$ nm from the gold surface. }
    \label{setup}
    \vspace{-5mm}
\end{figure}

The experimental setup is shown in Fig. \ref{setup}a. We optically trap silica nanospheres of diameter $177 \pm 9$ nm using a 1596 nm collimated laser beam with power $P=220$ mW that is directed into a vacuum chamber and focused using an in-vacuum aspheric lens with a numerical aperture of 0.5.  The nanosphere is loaded into the optical trap by shaking a glass substrate with deposited nanospheres above the trap region using a home-built PZT-based driver \cite{launcher}, and the particle is initially trapped at a pressure of $8-13$ Torr at room temperature taking advantage of damping from the background gas. The vacuum chamber can then be pumped down to a pressure of approximately $10^{-4}$ Torr while maintaining a stable trap or to high vacuum ($<10^{-5}$ Torr) with the addition of parametric laser feedback cooling.

A dielectric sphere of volume $V=4/3\pi a^3$ and index of refraction $n$ in vacuum experiences a trapping potential created by a single focused laser beam propagating in the $\vec{z}$ direction, with a gradient force given by \cite{harada_radiation_1996}
    
    \begin{equation}
	    \vec{F}_{grad}(\vec{r},\vec{z}) = - \frac{3 V}{2c} \left( \frac{n^2-1}{n^2+2} \right) \vec{\nabla} I(\vec{r},\vec{z}),
        \label{bead trap potential volume equation}
    \end{equation}
    where $n$ is the index of refraction, $c$ is the speed of light, and $I$ the intensity of the laser.  In the Rayleigh regime, the scattering force 
    \begin{equation}
     \vec{F}_{\mathrm{scat}}=\frac{128 \pi^5 a^6}{3 c \lambda^4}(\frac{n^2-1}{n^2+2})^2 I(r,z)\hat{z}
    \end{equation}
    slightly displaces the equilibrium position of the nanosphere of radius $a$ along the beam axis by approximately $1.5$ $\mu$m for our typical beam and nanosphere parameters. The trap frequency in the radial i.e. transverse to the direction of propagation of the laser and axial directions is approximately given by $\omega_r=\sqrt[]{\frac{6 I_0 w^2_0}{c \rho w^4} \left( \frac{n^2-1}{n^2+2} \right) },$
and $\omega_z=\sqrt[]{\frac{3 I_0 \lambda^2}{c \pi^2 \rho w^4_0} \left( \frac{n^2-1}{n^2+2} \right) }$, respectively, where $\rho$ is the density of the sphere, $\lambda$ is the wavelength, $I_0$ is the peak intensity, $w_0$ is the waist of the laser beam, and $w \approx w_0$ is the beam width at the trap location.
    
    At the focal point, the laser has a measured waist of $1.6\pm0.1$ $\mu$m, creating a trap with a potential depth of approximately 7000 K for a silica sphere with a diameter of 177 nm. The light scattered by the nanosphere is collimated by a lens with a numerical aperture of 0.25, and imaged onto a segmented InGaAs photodetector. As observed in Fig. \ref{setup}b(lower), the trapped sphere oscillates in the trap with a frequency of approximately $30$ kHz in the radial directions, and a frequency of $5$ kHz in the axial direction, in good agreement with theoretical expectations. Two radial (transverse) peaks are visible since the imaging axis is not perfectly orthogonal to the trap normal modes of oscillation, and the degeneracy in the transverse peaks is slightly broken. This can be attributed to astigmatism in the laser focus and by the preferred direction determined by the linear polarization of the trapping laser as observed in other single beam tweezer traps  \cite{novotny2012}.
    
\begin{figure}
	\begin{center}
		\includegraphics[width=1.0\columnwidth]{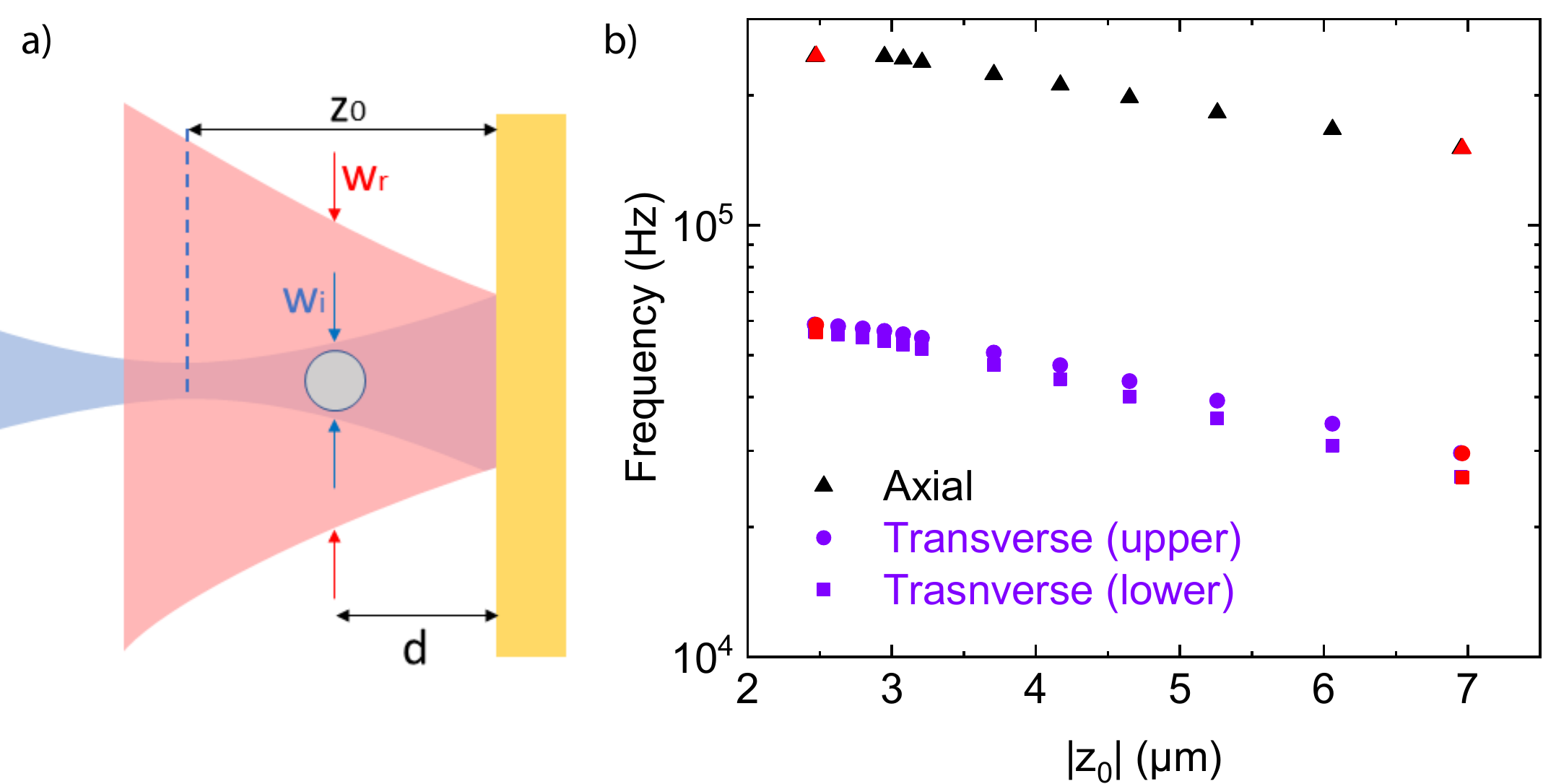}
	\end{center}
    \caption{(a) Illustration of beam parameters and distance parameters $z_0$ and $d$ defined in the text. (b) Transverse and axial trapping frequencies versus distance $|z_0|$ of beam focus to the mirror surface for a $177$ nm diameter silica sphere held at the first trapping antinode approximately at $d=408$ nm from the gold mirror. Data shown is recorded for the same nanosphere as the beam focal position is adjusted \textit{in-situ}. Red points indicate measured frequencies after returning back and forth to the same $|z_0|$.}
    \label{fig:lens}
\end{figure}

A microfabricated gold-coated silicon retro-reflector mirror of width $400$ $\mu$m and $100$ $\mu$m thickness is mounted on a piezo-driven in-vacuum translation stage which can move the mirror in a direction perpendicular to the laser beam's propagation at a short distance behind the trapped sphere. The mirror is moved into the laser beam to reflect the laser back onto itself creating a standing wave (Fig. \ref{setup}a). The nanosphere is robustly transferred to an anti-node of the standing wave which provides a localized confinement near the surface. By adjusting the distance $|z_0|$ between the laser beam focus and the mirror surface, the distance between the trapped nanosphere and the mirror surface can be controlled, up to a set of discrete available trapping anti-nodes each separated by one half of the laser wavelength. In practice, by adjusting the divergence of the input trapping beam using an adjustable telescope, the separation of the particle from the surface can be controlled at the $\sim \mu$m level.  The intensity profile of the Gaussian standing wave beam  \cite{standingwave}
in the paraxial approximation is \begin{eqnarray*}
   I(z,r)=I_0 \frac{w_0^2}{w_i^2}(e^{-2r^2/w_i^2}&+& \\ 2\rho_r\frac{w_i}{w_r}e^{-r^2/w_i^2}e^{-r^2/w_r^2}\cos{\phi} &+&\rho_r^2\frac{w_i^2}{w_r^2}e^{-2r^2/w_r^2}),
\end{eqnarray*}
where $w_0$ is the minimum waist of the incident Gaussian beam located at coordinate $z_0$, $w_i(z)=w_0\sqrt{1+\frac{(z+z_0)^2}{z_R^2}}$ is the $1/e^2$ radius of the incident beam, and $w_r(z)=w_0\sqrt{1+\frac{(z-z_0)^2}{z_R^2}}$ is the $1/e^2$ radius of the reflected beam, and the Rayleigh range $z_R=\pi w_0^2 / \lambda$. In our case where the focus occurs prior to striking the mirror surface, $z_0$ is negative. Beam parameters and distance parameters are illustrated in Fig. \ref{fig:lens}a. The peak intensity of the incident beam is $I_0=2P/\pi w_0^2$. The Fresnel reflection coefficient $r$ of the surface $r=\rho_r \exp{[-i\psi]}$ where $\psi$ is the phase shift acquired upon reflection from the surface and $\rho_r$ is the magnitude. The interference phase $\phi$ depends on the curvature $R_i(z)=-(z+z_0)[1+z_r^2/(z+z_0)^2]$ and $R_r(z)=(z-z_0)[1+z_r^2/(z-z_0)^2]$ of the incident and reflected wavefronts, respectively, as 
\begin{equation*}
    \phi=2kz-\frac{kr^2}{2}(\frac{1}{R_i}-\frac{1}{R_r})-\tan^{-1}{\frac{z+z_0}{z_R}}-\tan^{-1}{\frac{z-z_0}{z_R}}+\psi.
\end{equation*}
For a beam waist close to the mirror $z<<z_R$, we can approximate $w_i=w_r=w=w_0\sqrt{1+z^2/z_R^2}$ and the intensity profile is approximately 
\begin{equation}
    I(z,r) \approx I_0\frac{w_0^2}{w^2}e^{-2r^2/w^2}[1+2\rho_r\cos{(\phi)}+\rho_r^2],
\end{equation}
indicating a node at the metallic mirror surface for $\phi=\psi=\pi$, and antinode when $\cos{(\phi)}=1$.

Going beyond the paraxial approximation at 1st or 2nd order following the method of Ref. \cite{nonparaxial}, we find only slight changes in the expected trap frequency at the one percent level, for distances $|z-z_0|<z_R$. For a weakly focused beam with $w_0>>\lambda$, the first anti-node occurs at a distance of $\lambda/4$ from the conducting mirror surface, which corresponds to $399$ nm for a $1596$ nm laser. For our more tightly focused beam, the first trapping site occurs at approximately $408$ nm from the surface, considering the effects of wavefront curvature and the Gouy phase.  Fig. \ref{setup}b(upper) shows the position spectral density for a nanoparticle that has been transferred from the single-beam tweezer into the first anti-node from the mirror surface. Here the frequency of the radial motion has nearly doubled due to the nearly four-fold intensity increase in the standing wave trap at the anti-node, and the frequency of the axial motion has increased from $5$ kHz to approximately $225$ kHz due to the higher intensity and steep potential of the optical lattice along the laser propagation direction. 

It is also possible to load the particle into the second, third, or farther antinodes from the surface by adjusting the relative value of $z_0$ with the adjustable telescope, before slicing the retro-reflector mirror into the beam. In general it is possible to robustly retract the retro-reflector mirror, restoring the trapping potential to the single-beam tweezer configuration \textit{in-situ}, and back again into the standing wave trap without loss of the trapped particle. This allows for a dynamic adjustment of the position of the distance $d$ between the nanoparticle and mirror surface, by adjusting the initial focal position $z_0$ when the mirror is retracted.

When the particle is trapped at a particular antinode close to the surface, moving the focal position $z_0$ \textit{in-situ} by use of the adjustable telescope results in a change in the beam width at the location where the particle is trapped. The nanosphere tends to remain in the particular anti-node of the standing wave, while the beam width is tuned, resulting in a variation of the intensity and thus the trapping frequencies. Fig.\ref{fig:lens} illustrates the effect of tuning the telescope lens position \textit{in-situ} while the particle remains trapped at the first anti-node from the mirror surface. Upon moving the location of the waist $z_0$ in either direction, the frequency changes repeatably and upon returning to the same lens position matches the same trap frequency, suggesting the particle remains trapped at the same distance $d$ from the mirror surface throughout this process.

\section{Scanning force sensing} \label{sec:results}

\begin{figure}
	\begin{center}
		\includegraphics[width=1.0\columnwidth]{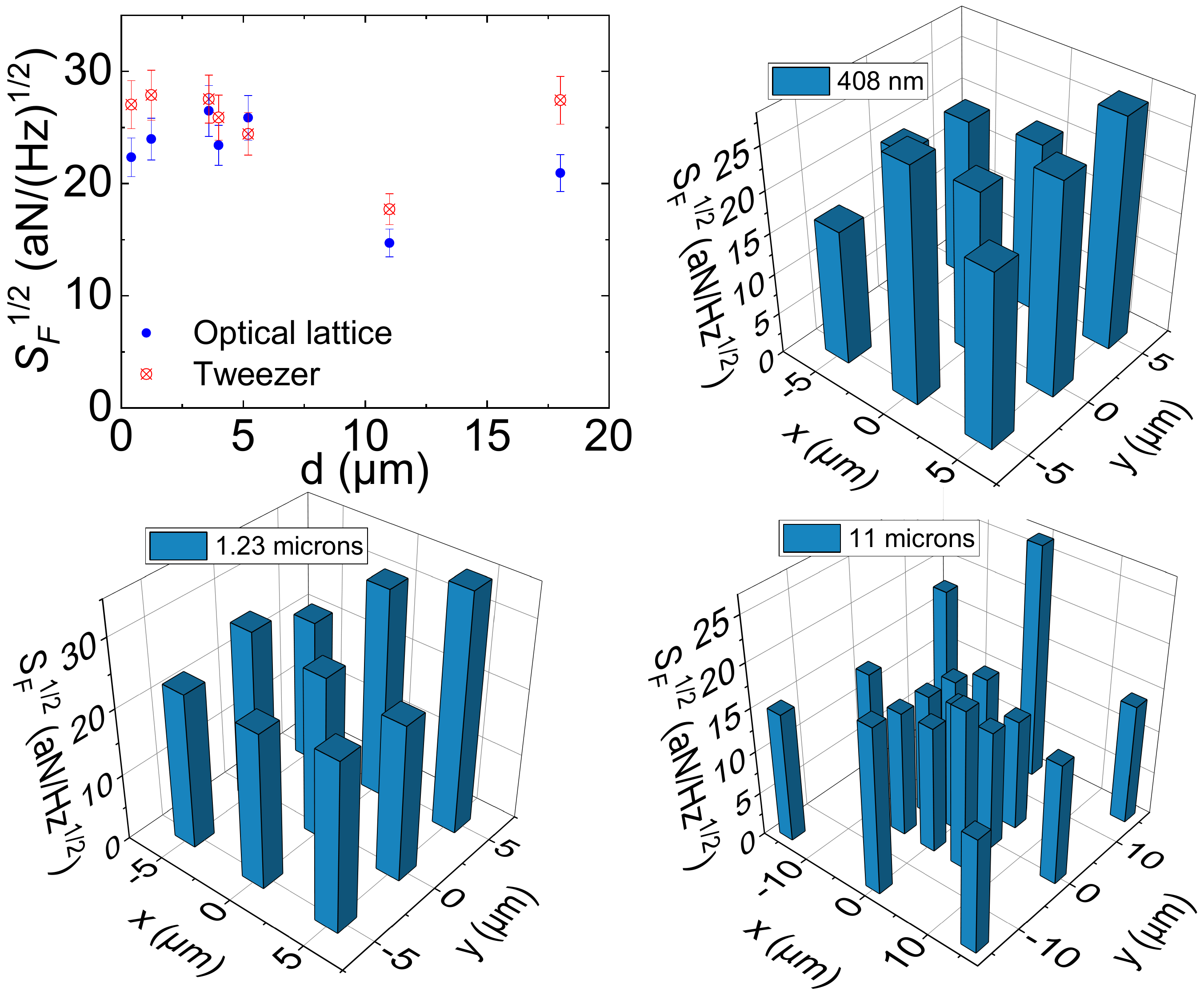}
	\end{center}
    \caption[frequencyandforce]{(upper left) Force sensitivity versus distance $d$ to the mirror surface for a $177$ nm diameter silica sphere in vacuum of $\approx 100$ mT at room temperature. Data is shown for three different nanospheres. The two points taken closest to the surface correspond to the same trapped particle. The data point at $\approx 11$ $\mu$m corresponds to a different particle, and the remaining points are all taken with a third trapped particle. A corresponding sensitivity measurement for the optical tweezer configuration with the retro-mirror retracted is also shown for each data point. (upper right, lower) Force sensitivity obtained while scanning the position of the particle in the plane parallel to the surface at distances of $d=0.408$, $1.23$, and $11$ $\mu$m, respectively.}
    \label{Fig:scanfig}
\end{figure}

We study the dependence of the force sensitivity of the detector as a function of distance $d$ from the gold mirror surface in Fig. \ref{Fig:scanfig}. The distance from the mirror surface is estimated from the observed trapping frequencies, taking into consideration the original single-beam trapping frequencies for each particular particle and the ratio of the standing wave trap frequencies to the single beam frequencies as the retro-reflecting mirror is introduced into the beam.  The force sensitivity and its uncertainty is determined from a calibration of the thermal noise spectrum taken at higher pressure of a few Torr where the nanosphere is in thermal equilibrium with the background gas, using the method reported in Ref. \cite{Atherton}.
The thermal-noise limited force sensitivity is 
\begin{equation}
    S_F^{1/2}=\sqrt{4 k_B T_{CM} m \Gamma}
\end{equation}
where $\Gamma$ is the damping coefficient and $T_{CM}$ is the temperature of the center-of-mass motion of the trapped particle of mass $m$. In addition to thermal noise, the effective temperature can be affected by technical noise, e.g. laser noise or other heating mechanisms such as radiometric forces \cite{Atherton} and photon recoil heating \cite{jain2016}. The minimum detectable force $F_{\rm{min}}=\sqrt{4 k_B T_{CM} m \Gamma / \tau}$ for averaging time $\tau$.
Each data point corresponds to 20 seconds of averaging time, and the minimum detectable force scales as expected with averaging time as $\tau^{-1/2}$ over this time interval. The dominant contribution to the experimental uncertainty comes from the error on the mass of the particles, which we estimate as approximately $15\%$, followed by uncertainty in the mechanical quality factor as determined from the peak width on the position spectral density. At the level of vacuum studied thus far, we see no significant degradation of force sensitivity near the surface, corresponding to force sensitivity at the level of a few attonewtons. The sensitivity agrees with expectations due to background gas pressure down to approximately 500 mT. Below this pressure we see effects of heating likely from radiometric forces, as observed in prior work \cite{Atherton}. We expect this sensitivity can be significantly improved by operation in the high vacuum regime, facilitated by use of three-dimensional parametric feedback cooling \cite{novotny2012} or cold-damping \cite{Ranjit_zeptonewton_2016}. Even at the moderate level of vacuum studied, the capability for scanning force microscopy with attonewton sensitivity at sub-$\mu$m distances from the surface is already realized, greatly exceeding the sensitivity of a typical scanning atomic force microscope \cite{saridreview}.

We are able to scan the position of the trap in the plane parallel to the surface at distances $d$ ranging from $\sim 0.4-18$ $\mu$m. The scanning is accomplished by an in-vacuum PZT-stage which changes the angle of the beam entering the trapping aspherical lens. The distance scale in the transverse ($x$-$y$) plane is calibrated by moving the quadrant photodetector on a two-dimensional micrometer-driven translation stage and considering the magnification of the optical imaging setup. Trapping at larger separation distances is also possible but is observed to occasionally result in the particle ``hopping'' between adjacent potential wells, as the relative intensity of the retro-reflected beam becomes diminished, resulting in less constructive interference and a shallower optical lattice along the beam axis. 

In Fig. \ref{Fig:scanfig} we show the force sensitivity map obtained by scanning over a grid of approximately $30 \times 30$ $\mu$m$^2$ at $d=11$ $\mu$m and of $10 \times 10$ $\mu$m$^2$ at $d=408$ nm and $d=1.23$ $\mu$m. Each data point corresponds to 20 seconds of averaging time. We see no significant variation in the force sensitivity with distance to the surface. There are differences at the $15-30$ percent level across the surface. These variances are largely within the experimental uncertainty of the force sensitivity.  This scanning approach could be used to study and determine local sources of noise and systematic effects associated with patch potentials, surface adsorbates, and other local surface phenomena in measurements of the Casimir force or exotic surface-force tests.  Operation in higher vacuum could enable similar studies with zeptonewton sensivitity \cite{Ranjit_zeptonewton_2016}.
    
\section{Discussion} \label{sec:discuss}
 
 We have demonstrated a method to maneuver a silica nanosphere in proximity to a planar conducting surface using a retro-reflected focused laser trap in vacuum. The position of the particle can be repeatably scanned by tens of microns in a plane parallel to the surface and the distance from the surface can be controlled in the $\mu$m-range by loading the particle into particular anti-nodes of the trapping field.  This opens up the possibility of attonewton level scanning force microscopy, which could lead to a number of applications including tests of surface forces such as the Casimir effect and searches for novel micron-range fifth forces predicted by several theories of physics beyond the standard model \cite{geraci_shortRange_2010}. By functionalizing the particle with a magnetic moment or electric charge, scanning magnetic resonance force microscopy and scanning potentiometry could be enabled by this approach.
 
 By implementing parametric feedback cooling method in three dimensions, as has already been done for other single-beam tweezer traps \cite{novotny2012, Rodenburg_quantumModel_2016}, scanning force sensing in the ultra-high vacuum regime could be enabled. This is promising for achieving zeptonewton level force sensitivity, which has already been demonstrated in other  (non-retro-reflected) standing-wave optical traps \cite{Ranjit_zeptonewton_2016}. To demonstrate the compatibility of this trapping approach with parametric feedback cooling, we also performed preliminary tests of parametric feedback cooling on the vertical and axial motion of the particle in the trap both when the retro-reflecting mirror is in and out of the beam. The parametric feedback cooling is achieved by measuring the product of the position and instantaneous velocity of the nanosphere and modulating the trapping intensity (and trap stiffness) based on the measurement via an Electro-optic modulator. The feedback signal is generated using a dual-channel low-noise lock-in amplifier (Zurich instruments HF2LI).  Implementing two-dimensional parametric feedback cooling to temperatures below 100 Kelvin, achieved with a cooling rate of approximately 200 Hz, permits operating a stable trap in the regime of high vacuum ($<10^{-5}$ Torr) and future work will include investigating scanning force sensing with three-dimensional feedback cooling.

\section{Acknowledgements} \label{sec:acknow}
    We thank J. Lim, E. Weisman, G. Winstone, and G. Ranjit for helpful discussions and assistance with electronic components.  This work is partially supported by NSF grant no. PHY-1806686, the Heising-Simons Foundation, and the office of Naval Research grant no.417315//N00014-18-1-2370.

\bibliographystyle{apsrev4-2}
\bibliography{References}

\end{document}